Great Chiral Fluorescence from Optical Duality Silver Nanostructures Enabled by 3D Laser Printing

Hongjing Wen[1,2,3], Shichao Song[1,3], Fei Xie[1], Bin Wang[2], Jian Xu[1], Ziwei Feng[1], Shiyu Wu[1], Jing Han[1], Baiou Guan[1], Xiaoxuan Xu[2*], Yaoyu Cao[1*], Xiangping Li[1]

[1]Guangdong Provincial Key Laboratory of Optical Fiber Sensing and Communications, Institute of Photonics Technology, Jinan University, Guangzhou 511443, P. R. China.
[2]The Key Laboratory of Weak-Light Nonlinear Photonics, Ministry of Education, School of Physics, Nankai University, Tianjin 300071, China.
[3]These authors contributed equally: Hongjing Wen, Shichao Song.
*email: yaoyucao@jnu.edu.cn; xuxx@nankai.edu.cn.

**Abstract**
Featured by prominent flexibility and fidelity in producing sophisticated stereoscopic structures transdimensionally, three-dimensional (3D) laser printing technique has vastly extended the toolkit for delivering diverse functional devices. Yet chiral nanoemitters heavily resorting to artificial structures that manifest efficient emission and tightly confined light-mater interactions simultaneously remains alluring but dauntingly challenging for this technique at this moment. In this work, we assert the chiral photoluminescence is implemented from silver nanostructures of optical duality in one go via a twofold three-dimensional laser printing scheme. Such laser printing protocol allows the highly desired duality by simultaneously producing uniformly distributed fluorescent silver nanoclusters and aggregated plasmonic silver nanoparticles to tightly confine chiral interactions at the nanoscale. A helical emitter of 550 nm-helix-diameter as fabricated has seen a record-high luminescence anisotropic factor with the absolute value up to 0.58, which is two orders of magnitude greater than fluorescent chiral silver clusters. This method holds great promise for future versatile applications in chiroptical nanodevices.

Chiral nanoemitters currently have attracted drastically increasing research interests for their abilities to underpin cutting-edge applications spanning the advanced optical and photonics[1-4], such as manipulating energy transfer among topologically configured photonic systems[5] and controlling communication with spin information[6]. Such functionalities and applications intensely rely upon implementing strong light-matter interactions in the nanoscale[7], which requires extremely confined interacting space to achieve demanding chiral luminescence that is not always applicable to the natural chiral materials with a common anisotropic factor at the level of 0.001[8]. Consequently, the advance of such chiral light sources preferably seeks artificial materials generated through deterministic nanofabrication technique with considerable printing freedom in favor of offering efficient emitters and tight light-matter interaction confinement at the same time.

3D laser printing, as one of the most important and fascinating technological innovations[9-12], has proven to be an indispensable micro/nanofabrication technique to shape multiple materials by employing their intrinsic optical nonlinear responses[13-17], such as multi-photon photoredox (MPP)[16], two-photon polymerization (TPP)[18-20]. During the past few decades, the printing strategy that achieves ultrahigh spatial resolution[21] has been used to additively create a myriad of functionalized structures for advanced nanoscale photonic and optoelectronic applications[22,23]. The flexibilities and fidelities rooted in the technique add vast advantages to arbitrarily and precisely producing stereo architectures for spatially manipulating electromagnetic field[24]. Nevertheless, most of these achievements feature with producing structures based on materials with mono-characteristic supremely well[25]. The limitation arising from lack of multifunctional materials and the corresponding printing protocols greatly bars the efforts in the subwavelength scale for elaborately delivering chiral regulation of radiative electromagnetic waves over the short wavelength bands and exploring unparalleled optoelectronic properties.

In this article, we propose and demonstrate photoreduction-based additive laser printing scheme to categorically achieve chiral luminescence with 3D silver nanostructures of optical duality in one go within agarose hydrogel, where the anisotropic factor of the photoluminescence with absolute value can reach up to 0.58 for a single emitter. The silver nanoclusters with the average size amid the mesoscopic scale between isolated metal atom and metallic nanoparticles[26-28] can be precisely produced by laser induced photoreduction of silver ions for light emission with its spectrum consistent with previous reports[29-31], whilst the silver nanostructure composed of assembled silver nanoparticles can be simultaneously fabricated to offer a delicate platform for the helicity control in visible wavelength regions with surface plasmon effect[32,33]. As such, the twofold laser printing protocol can readily create nanoemitters with steered luminescence through programmably forming two types of silver nanofeatures at the same time. The feasibility of this printing scheme is strongly bolstered by the use of the agarose matrix that renders both of pore grids and electron donors for fixing the silver nanoemitters and assisting the silver photoreduction, respectively. By producing glowing silver nanoclusters and fabricating spiral-shape nanostructures synchronously, silver nanocluster-based nanostructures with regulated photoluminescence and chiral properties are realized with multiple complexities. With the aid of plasmonic helical structures, the photoemission process of silver nanoclusters can be deeply guided to follow the structural symmetries. The absolute value of the highest anisotropic factor can reach up to ~0.58 at 675 nm, manifesting extraordinary chiral photoluminescence characteristic with the value two orders of magnitude greater than fluorescent chiral silver clusters[34]. Moreover, the theoretical calculations confirm the observation of circular dichroism (CD) and anisotropic factor of the emitter which opens up exceptional perspectives for delivering versatile applications in chiroptical nanodevices.

**Results**

As an appealing platform for hosting the production of silver nanostructures, the water-soluble agarose offers a homogenous circumstance for the reactants to stably photosynthesize silver nanoclusters (SNCs) and silver nanoparticles (SNPs) in the interior of the gel. The size-tunable pore structures of agarose[35], conducive to the aggregation and formation of silver nanofeatures, could scaffold 3D consecutive silver nanoclusters-based structures in effect, as shown in Figs. 1(a-c) (The scanning electron microscopy image of agarose can be seen in Supplementary Fig. S1). Moreover, taking advantages of multiple hydroxy groups of electron donor ability linked to the scaffoldings, agarose can aid the generation of SNCs and SNPs through anionic surfactant-assisted redox reactions[16] superior to its hydrogel counterparts (Seen in Supplementary Fig. S2). Herein, SNCs and SNPs are simultaneously photoreduced from silver ions for light emission and light manipulation, respectively, as shown in Fig. 1d. Eventually, the stable silver nanostructures composed of varisized SNCs and SNPs will be obtained tracing the printing path.

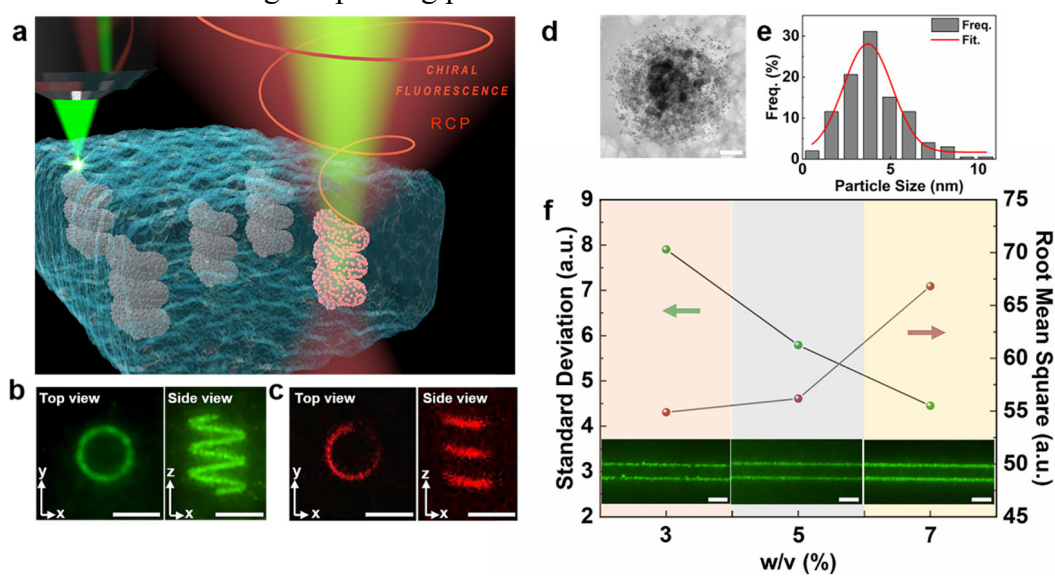

**Fig. 1. Stereo silver nanocluster-based structure via 3D laser printing. a** The schematic diagram of the formation of fluorescent silver helical structures in agarose matrix via 532 nm femtosecond laser printing and the chiral fluorescence generated from the silver structures excited by 532 nm continuous wave laser. **b** The dark field images of a 3-turn helical structure with the radius of 5 μm and the pitch distance of 5 μm in agarose matrix. The top view of the helix is on the left, while the side view is on the right. The scale bars are 10 μm. **c** The confocal fluorescence images corresponding to the 3D spiral structure of **b** in top view (left) and side view (right) at a certain focal plane. The scale bars are 10 μm. **d** TEM image of a single silver dot composed of SNCs (with sizes below 10 nm) and SNPs (with sizes over 10 nm) with different sizes in agarose. The scale bar is 100 nm. **e** The size statistical distribution of SNCs inside the silver dotted structure in **d**. The nano-entities with intense fluorescent effect below 2 nm account for ~20% out of the total SNCs. **f** The standard deviation (SD) and the root mean square (RMS) of the scattering intensity variance along the silver line fabricated in agarose with different mass-to-volume ratios (with details shown in Supplementary Fig. S3). Insets are the corresponding dark field images of the lines as printed. The scale bars are 10 μm.

The size distribution of the SNCs and SNPs is gauged via transmission electron microscope (TEM) image of a silver dot fabricated in agarose matrix with the laser power of 1 mW and the exposure time of 80 ms. SNCs below 2 nm with relatively superior luminescence ability[28] accounts for about 20% of nanofeatures below 10 nm where the statistical average size is at about 4 nm, as depicted in Figs. 1(d and e). A lump of assembled SNPs with the size of around 300 nm can be obtained in the central region, as a result of intense light field boosted series of chemical and physical reactions including photoreduction, trapping and coalescence. Meanwhile, few-nanometer SNCs can be found either attaching on the surface of a few larger SNPs like a shell or scattering nearby the core, which may give rise to a hybrid optical property. To improve the uniformity of the assembled SNPs, we investigated the light scattering intensity distribution of the silver lines as fabricated with varying agarose contents in terms of mass-volume ratio. As shown in Fig. 1f, an inverse correlation between the scattering intensity variance and the scaffolding density can be found with the standard deviation and the root mean square analysis. Nevertheless, while the content of agarose exceeded 7% w/v, the silver ammonia solution exhibited difficulties in spreading into the agarose matrix. Thus, we opted for the mass-volume ratio of 7% for the scaffolding matrix in the subsequent experiments. (The detailed data is shown in Supplementary Fig. S3) Through computer aided laser scanning, a rigid three-dimensional luminescent silver nanocluster-based helical structure can be successfully fabricated within the three-dimensional pore grids of agarose matrix. Figs. 1(b and c) clearly display the rigid scattering and luminescence patterns arising from the excitation of SNPs and SNCs, respectively, which refers to the surface plasmon effect and the irradiation process accordingly.

To investigate the optical irradiation properties of the luminescent silver structures, silver dots composed of SNCs and SNPs as elementary building blocks are adopted. As shown in Fig. 2a, the luminescence and scattering intensities of a dotted structure produced with the 532 nm-fs laser beam of power ($P_{lp}$) 1 mW is measured in situ with the spectral range of 600 nm to 800 nm (red line). The emission lifetime of the silver dot is gauged below 0.5 ns that implies a fast luminescence process (Supplementary Fig. S4a). The bright fluorescence occurring in the whole printing region reveals that the SNCs spread over the silver dot as shown in the fluorescent image captured via the confocal microscope. In the meantime, the scattering spectrum of the silver dot shows the light can be strongly scattered in the same range as a result of surface plasmon effect. The measured scattering peak virtually coincides with the simulation result, where the broader scattering band in the experimental spectrum can be attributed to the effective silver dot composed of varisized SNPs that can be seen in Supplementary Figs. S4(b and c). Moreover, the polarization of fluorescence from the silver dot is also studied. A linear polarizer (LP) is added into the fluorescence-collecting system to manifest the polarization characteristic of emission through changing the analyzer angles[36]. To avoid the disturbance from the analyzing system, the Rhodamine 6G (R6G) with unpolarized fluorescence is introduced into the agarose matrix system as a reference. Consequently, the relative intensity variation of fluorescence ($V_f$) emitted from the silver dot with the analyzer rotation angle θ can be defined as $V_f = (I_{dot}(\theta) - I_{R6G}(\theta))/(I_{dot}(0°) - I_{R6G}(0°))$, where $I_{dot}(0°)$ and $I_{R6G}(0°)$ are the fluorescence intensity of the dot and R6G reference with the initial polarization analyzer angle (0°), respectively. $I_{dot}(\theta)$ and $I_{R6G}(\theta)$ denote the fluorescence intensity of the dot and R6G reference with a rotated angle $\theta$, respectively. As depicted in Fig. 2b, the virtually unchanged $V_f$ value at around 1 measured at ~630 nm represents the structure possesses an unpolarized electromagnetic response, indicating the silver nanocluster fluorescence of non-polarizing characteristic (The measured fluorescent spectra with different polarization angle $\theta$ can be seen in Supplementary Fig. S5).

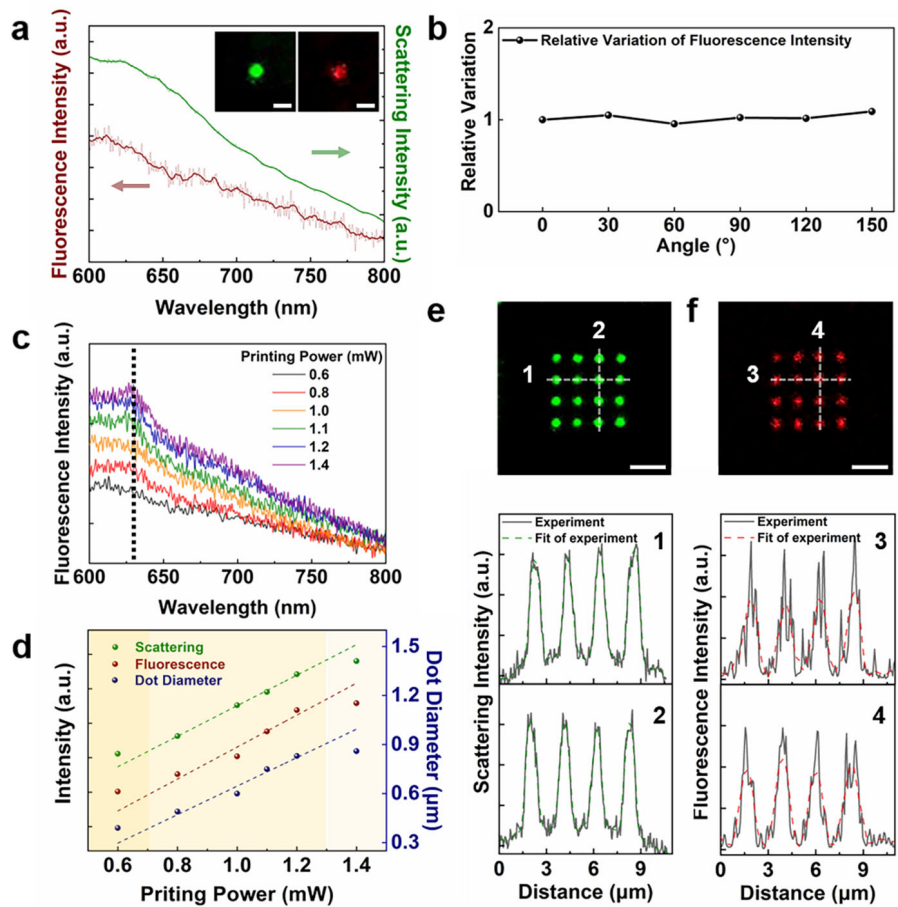

**Fig. 2. Optical duality of silver nanopatterns. a** The scattering spectrum and fluorescence spectrum excited by 532 nm CW laser of a silver dotted structure. The insets depict the dark-field image and the confocal fluorescence microscope image of the silver dotted structure fabricated in agarose with the laser power ($P_{lp}$) of 1 mW and the exposure time of 80 ms. The scale bars are 2 μm. **b** The relative variation $V$ of fluorescence intensity emitted from silver dotted with different polarization angles. **c** The fluorescence emission spectra of silver nanodots printed with different $P_{lp}$ (from 0.6 to 1.4 mW). **d** The corresponding fluorescence intensity (red ball scatters) at the wavelength of 630 nm (black dot line in **c**), the scattering intensity (green ball scatters) and dot diameter (blue ball scatters) under different printing powers. **e** The dark field image of 4×4 silver dot lattice with the printing power of 1 mW and exposure time of 80 ms for each dot. And the corresponding scattering intensities along the two white dashed line marks are depicted the nether figures, which shows favorable uniformity of the lattice structure. The scale bar is 5 μm. **f** The fluorescence image of the 4×4 silver dot lattice. And the corresponding fluorescence intensities of the two white marked areas are depicted nether figures, which shows reasonable uniformity of photoluminescence. The scale bar is 5 μm.

To further explore the fluorescence and scattering intensity with regard to the size of a single pixel as fabricated, a series of dotted structures were realized with different laser power $P_{lp}$ ranging from 0.6 to 1.4 mW. And the feature sizes of the silver dotted structure measured via the dark-field images can be found in the range from 380 nm to 830 nm. In effect, the fluorescence intensities are positively correlated with the dimensions of silver dots, as shown in Fig. 2c. In the meantime, we can find the variance of scattering intensity follows a similar tendency, indicating a reinforced ability of light manipulation with increasing the printing power, where the measured dark-field images of silver dotted structures corresponding to fluorescence spectra are presented in Supplementary Fig. S6a. As shown in Fig. 2d, a virtually linear relevance between the fluorescence intensity, the scattering intensity and the silver dot size can be found with increasing the printing power from 0.8 mW to 1.2 mW, revealing a single photon characteristic for the generation of silver dots as well as the corresponding optical properties in the particular printing power region. Fluorescence emission spectra of a silver nanodot in agarose matrix obtained with CW laser excitation can be seen in Supplementary Figs. S6(b-c). In addition, the accumulation rate of the dot size and the related optical responses can be found slowing down with further increasing printing laser power, which may arise from the depletion of the silver salts in the fabrication region.

In order to investigate the uniformity of the fluorescence and the surface plasmon based scattering from silver structures, we fabricated a 4×4 silver dot lattice with the laser power of 1 mW and the exposure time of 80 ms. The corresponding dark field image (Fig. 2e) and fluorescent image (Fig. 2f) measured with confocal fluorescence microscopy reveal a homogenous feature for the dot pattern. As it mainly comes from the SNPs, the steady high scattering intensity from dot to dot reflects the glorious uniformity between each dot unit as well as a powerful light manipulation ability of assembled SNPs as produced resorting to the surface plasmon effect, which can be obtained through the dashed mark line 1 and 2. Additionally, the uniformity of fluorescence is estimated through the dashed mark line 3 and 4, and emission intensity fluctuations are less than 10%, which further confirms the favorable homogeneity of SNCs as produced simultaneously.

Harnessing multiphoton absorption characteristic, silver ions in the agarose hydrogel are photoreduced to form spatially resolved arbitrary silver structures by the computer aided printing with assembled SNPs and SNCs for effective light manipulation and fluorescence, respectively. In the meantime, the fluorescence intensity of the demonstrated structures can also be regulated by adjusting fabrication parameters such as the laser intensity. As shown in Fig. 3a, dark field image (left upper green graph) and confocal fluorescence microscope image (left nether red graph) of the 2D dolphin pattern confirm the feasibility of the steady patterning. Besides, the right side shows the variation of scattering intensity and fluorescence intensity form the patterns depending on the printing power corresponding to the subregions in left side. In this case, the fluorescence of the corresponding pattern was excited by 532 nm CW laser with the excitation power of 200 µW so as to ensure that the fluorescence will not be ablated. With a multilayer point-by-point printing approach, a homogeneous layered structure with interlayer spacing of 5 μm can be explicitly fabricated and resolved in both of dark-field images and confocal fluorescence microscope images, presented in Fig. 3b. Furthermore, by implementing 3D vector scanning scheme, a cubic scaffold with continuous sides can be readily achieved Fig. 3c. The corresponding dark-field images in the perspective view and top view indicate the high continuity of the stereo structure as a result of accurately assembling SNPs as supposed. The confocal images in perspective view are acquired by stitching layer records (The detailed layer-by-layer scanning images can be found in Supplementary Fig. S7). While a bit fluctuation of confocal fluorescence intensity in the perspective view comes from the image mosaic errors during the multilayer scanning, the fluorescence intensity distribution in the top view reveals reasonable uniformity for the same layer. Herein, eight lines are marked in the top view of the confocal fluorescence image to verify the uniformity of the 3D continuous structures. As shown in Fig. 3d, the fluctuation of fluorescence peak intensity is less than 10%, indicating considerable continuity and uniformity of the stereo structure which is in favor of achieving versatile fluorescent nanostructures.

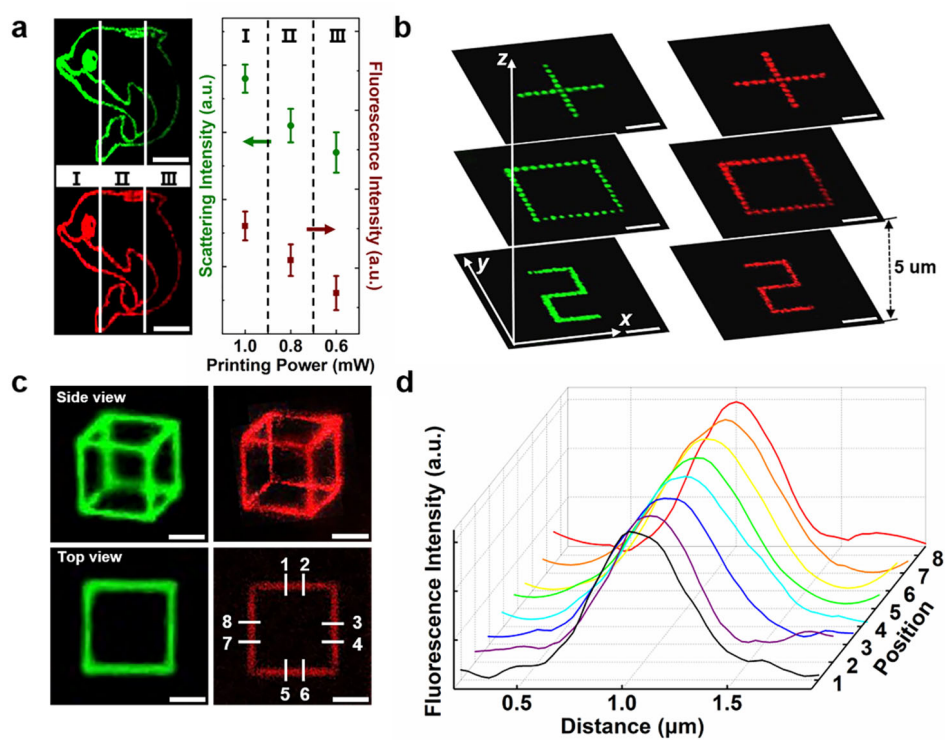

**Fig. 3. Structuring and patterning of 3D fluorescent silver nanoclusters-based structures. a** The dolphin pattern fabricated by three different printing powers (1.0 mW for the subregion I, 0.8 mW for the subregion II and 0.6 mW for the subregion III). In the left side, the left upper green graph (false-color image) is the dark field scattering image and the left nether red one is the confocal fluorescence image. The right side shows the variation of scattering intensity and fluorescence intensity form the patterns depending on the printing power corresponding to the subregions in left side. **b** The dark field (left) and fluorescence (right) images of three layers of silver patterns printed with the layer spacing of 5 μm. **c** The dark field (left) and fluorescence (right) images of 3D cubic structure fabricated in agarose matrix on side view and top view, respectively. **d** The fluorescence intensity distribution tracing the white short lines marked at eight positions on the top of cubic structure in c. All the scale bars are 5 μm.

As a representative three-dimensional structure with chirality, the helix possesses vast polarization manipulation capabilities for realizing circular dichroism (CD) and optical rotations[4,37]. Taking full advantages of the chirality of the helical structure and photoluminescence of SNCs, fluorescent chiral structures can be realized in agarose by utilizing the joint characteristics from the inherent structural uniqueness of SNP spiral lines and fluorescent spectrum attribute of SNCs. To characterize the chirality of the helical structure, CD is defined as the difference between the fluorescent intensity with left-handed circular polarization (LCP) and right-handed circular polarization (RCP), which can be calculated by using:

$$CD = T_{\text{LCP}}(\lambda) - T_{\text{RCP}}(\lambda) \tag{1}$$

where $\lambda$ is the light wavelength, $T$ denotes the transmitted fluorescence light intensity. And the anisotropic factor $g$ defined as the differential circular transmitted fluorescence intensity normalized by the average fluorescence intensity is given as the follow[38]:

$$g = \frac{2(T_{\text{LCP}}(\lambda) - T_{\text{RCP}}(\lambda))}{T_{\text{LCP}}(\lambda) + T_{\text{RCP}}(\lambda)} \tag{2}$$

As illustrated in Figs. 4(a-d), a 2-turn left-handed helical structure with the helix diameter of 550 nm and the pitch of 550 nm in z direction was fabricated with the laser power of 0.1 mW and the printing speed of 1 μm/s. The scattering images of the helical structure on the top view and side view are snatched by the confocal microscope, as presented in Figs. 4(a and c). The linewidth of the spiral line in the helical structure is measured by the full width at half maximum (FWHM) of the scattering intensity peak at ~200 nm in horizontal orientation and ~340 nm in longitudinal direction, shown in Figs. 4(b and d) (More helical structures with different parameters can be found in Supplementary Fig. S8).

The most prominent helicity characteristic for the helical structure relies on the repeated circles along the helical axial. By gradually changing the turns from n = 2 to 4, we can accordingly tune the structural chirality, where the corresponding CD spectra can be obtained by exciting the luminescence at the far-end of the emitters to the detector. As shown in Fig. 4e, the bands of the CD undergo manifest broadening over the wavelength measurement range of 630 nm to 750 nm with increasing the turns of the helical structure, while the maximum absolute value of CD increases gradually from 204 to 315. In addition, the fluorescence CD of the emitter clearly reveals a wavelength-dependence property. We can find the CD value of the helical structure could dwindle more than 50% as the fluorescence of the emitter drops in the long-wavelength region.

To investigate the chirality functioning mechanism of silver helical nanostructures as fabricated, the finite difference time domain (FDTD) method is utilized with the dipole sources as the radiative origins of fluorescence and the power profile monitors for collecting the energy flux through the helical structures, where the spans of the monitors are in line with the objective aperture used in experiment (More minutiae about the simulation settings are shown in Supplementary Fig. S9). Through elaborately building the silver structure models, the intensity of the transmitted LCP and RCP light can be extracted separately (The experiment and calculation results of LCP and RCP light are shown in Supplementary Fig. S10). As revealed in Fig. 4f, we can verify the broadening CD peak with increasing the helix turn, which may be attributed to the formation of multiple modes stemming from the structural plasmonic resonance[37]. Moreover, the calculations possess coordinating tendency sticking to the measured CD spectra which confirms that the fluorescent silver nanohelix can afford a favorable chiral fluorescence. While the simplified silver helical structure model used in the simulations with ideally smoothed outer surface renders a slightly flat spectral response, the varisized silver nanoparticles existing on the surface of helical structures in practice may cause diverse localized plasmonic resonances, resulting in fluctuated CD values.

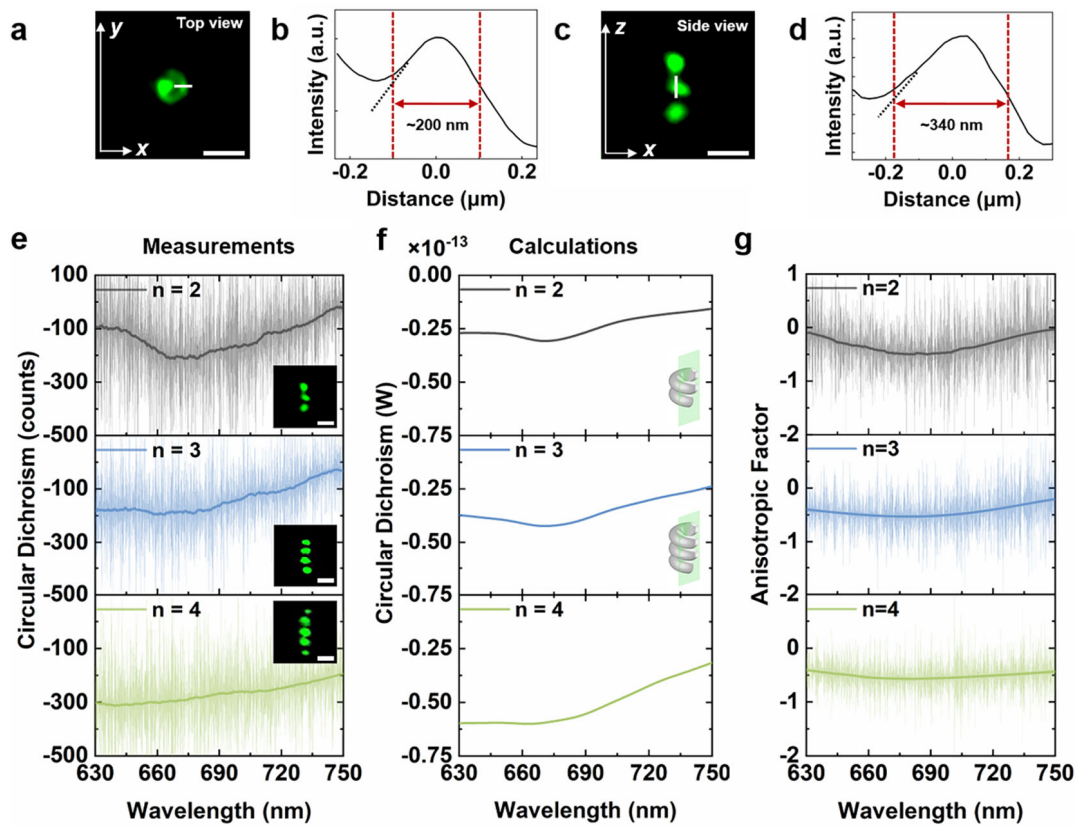

**Fig. 4. Chiral fluorescence of silver nanocluster-based helical nanostructures. a** The top-view scattering image of a silver nanocluster-based helical structure (n = 2) captured by the confocal microscope. **b** The linewidth of the spiral line for the helical structure (obtained at the FWHM) is measured as ~200 nm in the horizontal orientation. **c** The scattering image of the side view of the helical structure. **d** The linewidth of the spiral line for the helical structure is measured as ~340 nm in the longitudinal direction. **e** The experimental results of CD for single silver nanoemitters with the helix turns ranging from 2 to 4. The insets are the scattering images of the cross-section profiles. **f** The calculated CD values normalized to the emission intensity of the dipoles used in FDTD method (The inset is the simulation model set up with the same parameters obtained in experiments: helix diameter of 550 nm and pitch of 550 nm in *z* direction). The corresponding models in calculations are shown in insets, and the green plane represents the measured plane with respect to **e**. **g** The measured anisotropic factor spectra for single silver nanoemitters obtained with different turns of the helical structures. All the scale bars are 1 μm.

In order to further characterize the chirality of the silver nano-helix, the structural anisotropic factors of the fluorescence varying with wavelength can be obtained through gauging the transmittance of the LCP and RCP light with a microscope spectrometer, which is shown in Fig. 4g. The absolute values of anisotropic factor ranging from 0.05 to 0.58, which depends on either the structure characteristic or the emission wavelength, clearly reveal a prominent CD for the silver spiral nanoemitter. As the number of turns increases from 2 to 4, the absolute value of anisotropic factor becomes greater from the maximum of ~0.50@684 nm to ~0.58@675nm, respectively. In the meantime, we can find a decrease in the slope of the wavelength-dependent anisotropic factor curve with increasing the spiral turns, indicating a broadening band for yielding chiral fluorescence accordingly.

**Discussion**

Benefiting from the twofold laser printing method, 3D silver chiral nanoemitters consisted of plasmonic nanostructures for tightly confining chiral interactions and photoluminescent plasmonic nanoclusters can be precisely fabricated with tunable morphology and optical characteristics in one go. Along with the printing process, two types of silver nanofeatures are produced simultaneously in favor of stable photon emission and uniform light manipulation enabled by SNC-based fluorescence and SNP-based surface plasmon effect, respectively, which contributes to a scale-regulated joint function of optical duality. In the meantime, the size and optical properties including the intensity of scattering and luminescence of the nanostructures as fabricated can be accurately controlled with printing parameters. By elaborately designing and fabricating silver helical nanostructures for spatially steering light passing through, photoluminescence-regulatable emitters with appreciable circular dichroism can be readily realized. The value of the luminescence anisotropic factor can be greatly escalated compared to its chiral silver nanocluster counterparts. Besides, theoretical calculations of these distinctive helical structures are also investigated to the check for the consistency with experimentally measured results, revealing the feasibility in creating multiple functional nanodevices with extraordinary optical chirality manipulation. This technique allows intentionally and precisely exerting 3D chiral symmetries and orderings on nanostructures with active optical response for exploring far more morphological and structural complexity as well as advanced functionalities.

**Methods**

**Reagents and materials.** Materials for photoreduction of fluorescent silver nanoclusters/agarose matrix system are mainly consisted of agarose gel (Gene Company, LTD), diamine silver ions (DSI) aqueous solution as the silver resource and a nitrogen-atom-containing alkyl carboxylate (n-decanoylsarcosine sodium, NDSS) as the surfactant (Tokyo Chemical Industry). The chemical reagents, such as agarose, silver nitrate, ammonium solution and NDSS were commercially available in Sigma-Aldrich at least of analytical grade and used without furtherly purifying. All aqueous solutions were prepared using ultrapure water (Smart2Pure, ThermoFisher).

**Agarose gel**. 0.7 g of agarose powder (gel strength (1%) 750 gr/cm$^2$, gelling temperature 37±1.5 °C) and 10 mL ultrapure water were mixed in glass container and heated until a clear viscous agarose gel solution was formed. The hot gel was then spread in a petri dish for 10 min at 25 °C until the gel solidify completely. Approximately 1 mm-thick square gel used for the subsequent laser printing was obtained from the gel block using a sharp blade.

**DSI aqueous solution.** Silver nitrate (0.3514 g) was weighed into a glass container and an appropriate amount of ultrapure water (1.0343 mL) was added to obtain a 2 mol/L silver nitrate solution. Ammonium solution ($NH_3$: 25%) is gradually added to the silver nitrate solution by a pipette. When the mixed solution of silver nitrate and ammonia solution became clear and transparent (ammonia water needs ~550 μL), the dropwise addition was stopped to obtain the DSI aqueous solution with the silver ion concentration of 1.3057 mol/L. Then, the configured DSI aqueous solution was diluted to obtain the DSI aqueous solution with a concentration of 0.1 mol/L for experiment.

NDSS solution: 0.1 mol/L NDSS solution was prepared from 0.0717 g of NDSS solid powder (CAS: 30377-07-2 > 98.0%) and 2.7022 mL of ultrapure water.

**3D laser printing.** The printing setup was employed to produce the transdimensional structures in fluorescent silver nanoclusters/agarose matrix system. A femtosecond laser beam from a 532 nm-wavelength femtosecond laser with repetition of 80 MHz and pulse width of 210 fs was directly focused onto the sample mounted on a piezo multi-axis stage through an oil immersion objective lens (×100, NA = 1.4, Olympus). The size of the focused printing laser spot is estimated at ~232 nm, calculated by the formula, spot size D = 0.61 × $\lambda_{lp}$/NA, where $\lambda_{lp}$ is the wavelength of laser beam[39].

A CCD camera (MER-132-43U3M-L, Daheng) that provides a bright field microscope magnified field of view in visible light region was used to monitor the fabrication process of the structures.

**Measurement of fluorescence spectra.** The sample was immersed in deionized water to remove the excess silver ions and then placed on the piezo multi-axis stage to be precisely positioned via CCD camera. Therefore, the fluorescence emission spectra were recorded in situ from the silver nanostructures using the spectrometer (SR-ASZ-0103, Andor). By introducing the combination of a linear polarizer and an achromatic quarter-wave plate (200 nm-800 nm), the received fluorescence spectra can be separated into left-handed and right-handed circular polarization[40]. Herein, in order to distinguish the dark-field images, the fluorescence images were processed with pseudo color according to their raw values.

**Data availability**

All data needed to evaluate the conclusions in the paper are present in the paper and/or the Supplementary Materials. Additional data related to this paper may be requested from the corresponding authors upon reasonable request.

**Acknowledgments**

This research was supported by National Key R&D Program of China (Grant No. 2018YFB1107200), National Natural Science Foundation of China (NSFC) (Grant Nos. 61605061, 61875073), Guangdong Provincial Innovation and Entrepreneurship Project (Grant No. 2016ZT06D081), Natural Science Foundation of Guangdong Province (Grant No. 2016A030313088), Guangdong Basic and Applied Basic Research Foundation (Grant No. 2019A1515110385) and the Fundamental Research Funds for the Central Universities.

**Author contributions**
Y.C. conceived the project. Y.C. and X. L. supervised the project. H.W., S.S. and Y.C. designed all experiments. H.W. conducted the fabrication and characterization of the silver helix. S.S. conducted the simulations for obtaining the calculated spectra from the laser printing structures. F.X. supported the fabrication and characterization process. H.W., S.S. and Y.C. prepared the manuscript and figures, and all authors participated in experiment data analysis and manuscript editing.

**Competing interests**
The authors declare that they have no competing interests.